\documentclass[a4paper,11pt]{article}
\usepackage{graphicx,amssymb,amsmath,amsfonts,epsfig}
\usepackage[usenames,dvipsnames]{color}
\usepackage{subfigure}
\title{Energy Formula, Surface geometry and Energy Extraction for Kerr-Sen Black Hole}
\author{Parthapratim Pradhan\footnote{pppradhan77@gmail.com}\\ 
\textit{Department of Physics}\\
\textit{Hiralal Mazumdar Memorial College For Women}\\
{Dakshineswar, Kolkata-700035, India}}
\date{}

\begin{document}

\maketitle

\begin{abstract}
We evaluate the \emph{surface energy~(${\cal E}_{s}^{\pm}$), rotational energy~(${\cal E}_{r}^{\pm}$) 
and electromagnetic energy~(${\cal E}_{em}^{\pm}$)} for a \emph{Kerr-Sen black hole~(BH)} 
having the event horizon~(${\cal H}^{+}$) and  the Cauchy horizon~(${\cal H}^{-}$). 
Interestingly, we find that the \emph{sum of these three energies is equal to the
mass parameter i.e. 
${\cal E}_{s}^{\pm}+{\cal E}_{r}^{\pm}+{\cal E}_{em}^{\pm}={\cal M}$}. 
Moreover in terms of the \emph{ scale parameter ~$(\zeta_{\pm})$, the distortion 
parameter~($\xi_{\pm}$) and a new parameter~$(\sigma_{\pm})$} which corresponds to the 
area~(${\cal A}_{\pm}$), the angular momentum ~$(J)$ and the charge parameter~($Q$), 
we find that the  \emph{mass parameter in a compact form} 
${\cal E}_{s}^{\pm}+{\cal E}_{r}^{\pm}+{\cal E}_{em}^{\pm}={\cal M} =\frac{\zeta_{\pm} }{2}
\sqrt{\frac{1+2\,\sigma_{\pm}^2}{1-\xi_{\pm}^2}}$
which is valid {through all the horizons} (${\cal H}^{\pm}$). We also 
compute the \emph{equatorial circumference and polar 
circumference} which is a gross measure of the BH surface deformation. It is shown that when 
the spinning rate of the BH increases, the \emph{equatorial circumference increases} while 
the \emph{polar circumference decreases}. We show that there {exist} two 
classes of geometry separated by $\xi_{\pm}=\frac{1}{2}$ {in} 
Kerr-Sen BH. In the regime  $\frac{1}{2}<\xi_{\pm}\leq \frac{1}{\sqrt{2}}$, 
the Gaussian curvature is negative and there {exist} \emph{two polar caps} on 
the surface. While for 
$\xi_{\pm}<\frac{1}{2}$, the Gaussian curvature is positive and the surface will be an oblate 
deformed sphere. Furthermore, we  compute the exact expression of \emph{rotational energy that 
should be extracted  from the BH via Penrose process}. The maximum value of rotational energy 
which is extractable should occur for \emph{extremal Kerr-Sen BH} i.e. 
${\cal E}_{r}^{+}
=\left(\sqrt{2}-1\right)\sqrt{\frac{J}{2}}$. 

%

\end{abstract}


\section{Introduction}
One of the fascinating objects in Einstein's general theory of relativity is 
a black hole (BH). It is fascinating  in the sense that  it has entropy which is 
proportional to the BH surface area~\cite{bk73}~(a two-dimensional surface)  
and the temperature which is proportional to the surface gravity~\cite{bcw73}.

The surface geometry of Kerr-Newman class of solutions to Einstein's gravitational 
field equations was studied by Smarr~\cite{smarr73c} in detail but the surface geometry 
of \emph{Kerr-Sen BH} i.e. charged rotating BH in heterotic string theory is not being 
studied so far in the literature. We want to fill this gap in the literature. The fact is 
that string theory is a leading candidate to unify gravity with the rest of the fundamental 
forces in nature. Hence the study of BHs in string theory plays an important role in the 
current investigation.

Recently~\cite{pp21}, we have studied in detail the energy formula for the NUT class of BHs. 
Specifically, we computed the surface energy, rotational energy, and electromagnetic energy 
for all the horizons. Remarkably, we proved that the sum of these three energies is equal to the 
mass parameter. This was tested for Taub-NUT BH, Reissner-Nordstr\"{o}m-Taub-NUT BH, 
Kerr-Taub-NUT BH and Kerr-Newman-Taub-NUT BH. In each case, we showed that
the sum of these energies is equal to the mass parameter.

Inspired by the above work, in the present work we want to evaluate these three
energies \emph{for a charged spinning  BH in heterotic string theory} having the event horizon 
and the Cauchy horizon in contrast with the Kerr-Newman BH.  By computing this energy we 
prove that the sum of three energies is equal to the mass parameter. \emph{Additionally}, we show 
that in terms of the scale parameter~$(\zeta_{\pm})$, the distortion parameter~($\xi_{\pm}$) and a 
new parameter~$(\sigma_{\pm})$ which corresponds to the BH area~(${\cal A}_{\pm}$), the BH angular 
momentum ~$(J)$ and the BH charge parameter~($Q$) then the mass parameter should  be written in a 
most compact form as
\begin{eqnarray}
{\cal E}_{s}^{\pm}+{\cal E}_{r}^{\pm}+{\cal E}_{em}^{\pm} &=& {\cal M} 
=\frac{\zeta_{\pm} }{2} \sqrt{\frac{1+2\,\sigma_{\pm}^2}{1-\xi_{\pm}^2}} 
\end{eqnarray}
which is valid for both the horizons~(${\cal H}^{\pm}$). It should be noted that $M$, $J$ and $Q$ are  
\emph{conserved quantities in the Einstein frame}.

When material or any physical objects are in a spinning motion, their surfaces become deformed. This deformation 
could be measured by equatorial circumference and polar circumference. Invariant measures of the 
surface geometry are calculated by using Gaussian curvature. We want to calculate the 
\emph{proper equatorial circumference and the proper polar circumference} which is a gross measure 
of BH surface deformation. It has been shown that when the spinning rate of  BH increases, the 
equatorial circumference increases while the polar circumference decreases. We prove that there 
exist two classes of geometry separated by $\xi_{\pm}=\frac{1}{2}$ for Kerr-Sen BH. In the regime  
$\frac{1}{2}<\xi_{\pm}\leq \frac{1}{\sqrt{2}}$, the Gaussian curvature is negative, and there exist 
two polar caps on the surface. While for $\xi_{\pm}<\frac{1}{2}$, the Gaussian curvature is positive 
and the surface is likely to be an oblate deformed sphere.

Furthermore, we evaluate the rotational energy that should be extracted from the
BH via \emph{Penrose process}. 
The maximum value of extracted energy should be found for maximally Kerr-Sen BH i.e.
\begin{eqnarray}
{\cal E}_{r}^{+} &=& \left(\frac{\sqrt{2}-1}{2}\right)\sqrt{2{\cal M}^2-Q^2}
=\left(\sqrt{2}-1\right)\sqrt{\frac{J}{2}}
\end{eqnarray}
In the limit $Q=0$ or $J={\cal M}^2$  i.e. for extremal Kerr BH, the extractable rotational energy 
is $ {\cal E}_{r}^{+}=\left(1-\frac{1}{\sqrt{2}}\right){\cal M}$. 
To the best of our knowledge, this is the \emph{first time we have reported in the literature such   
a remarkable result for a charged rotating BH in heterotic string theory}. 
All the above results discussed so far have been derived strongly motivated by the work of 
Smarr~\cite{smarr73c} [See also~\cite{smarr73a,smarr73b}].

In the next section~(Sec. 2) we will familiarize ourselves with the Kerr-Sen metric and 
its thermodynamic properties. In Sec.~3, we will derive the surface energy, the rotational energy, 
and the electromagnetic energy for Kerr-Sen BH. In Sec.~4, we will introduce the scale parameter 
and the distortion parameter for Kerr-Sen BH. In Sec.~5, we will introduce the intrinsic surface 
geometry of Kerr-Sen BH. In Sec.~6,  we will derive the BH horizons, angular velocity,  
surface area, and irreducible mass in terms of scale parameter and distortion Parameter. 
In Sec.~7, we will compute the gross measure of the BH surface deformation in terms of 
the equatorial circumference and polar circumference. In Sec.~8, we will calculate the 
amount of rotational energy that should be extracted from 
the BH in heterotic string theory. Finally, in Sec.~9 we have given our conclusions. 
In Appendix-A, we will compute the Gaussian curvature for Kerr-Sen BH which measures the 
intrinsic property of surface geometry. In Appendix B, we will calculate the Euler number of the 
BH  using the Gauss-Bonnet theorem.

\section{\label{ksen}The Kerr-Sen BH metric and its properties:}
The metric of rotating charged BH solution~[in the geometrized units i.e. ($c= G = 1$)] 
in heterotic string theory~\cite{as92} is 
$$
ds^2 = -\frac{\Delta-a^2 \sin^2\theta}{\rho^2}\,dt^2-\frac{4a{\cal M}r\sin^2\theta}{\rho^2}\,dt\,d\phi +
\frac{\rho^2}{\Delta}\,dr^2
$$
\begin{eqnarray}
+\rho^2\, d\theta^2+ \frac{\Xi}{\rho^2} \sin^2\theta\, d\phi^2 ~\label{seneq}
\end{eqnarray}
where
\begin{eqnarray}
  \rho^2 &=& r^2+a^2\cos^2\theta+2qr \\
  \Delta &=& r^{2}-2({\cal M}-q)r+a^2 \\
\Xi &=& \left(r^2+a^2+2qr \right)^2-\Delta a^2\sin^2\theta\\
       q &=& \frac{Q^2}{2{\cal M}}
\end{eqnarray}
The above spacetime metric obtained in the Einstein frame implies that there exists a BH solution which 
consists of the \emph{conserved quantities} i.e. mass~(${\cal M}$), charge~($Q$), angular momentum ($J$). 

There exist two horizons for Kerr-Sen BH namely the event horizon~(${\cal H}^+$) or outer horizon
and Cauchy horizon~(${\cal H}^-$) or inner horizon. Therefore the horizon radius may be determined
from the following  functions:
\begin{eqnarray}
\Delta \equiv \Delta (r) = r^{2}-2\left({\cal M}-q\right)r+a^2=0  ~.\label{horizon}
\end{eqnarray}
which gives
\begin{eqnarray}
r_{+}&=& \left({\cal M}-q\right)\,\, + \sqrt{\left({\cal M}-q\right)^{2}-a^2}\\
r_{-}&=& \left({\cal M}-q\right)\,\, - \sqrt{\left({\cal M}-q\right)^{2}-a^2}
~.\label{hr}
\end{eqnarray}
Note that $r_{+}>r_{-}$. Where $r_{+}$ is the radius of outer horizon 
while $r_{-}$ is the radius of inner horizon.

From Eq.~(\ref{horizon}) one must see that the horizon disappears unless $a\leq\left({\cal M}-q\right)$.
Thus the extremal limit of the Sen BH corresponds to $a=\left({\cal M}-q\right)$.
and the horizon for extremal Sen BH is located at $r_{ex}=r_{+}=r_{-}=a=\left({\cal M}-q\right)$.
The ergosphere for Kerr-Sen BH is located at
\begin{eqnarray}
r_{ergo}(\theta) &=& \left({\cal M}-q\right)\,\, + \sqrt{\left({\cal M}-q\right)^{2}-a^2\cos^2\theta}
~.\label{ergo}
\end{eqnarray}
The thermodynamic quantites like the BH area is 
\begin{eqnarray}
{\cal A}_{\pm} &=& \int^{2\pi}_0\int^\pi_0  \sqrt{g_{\theta\theta}g_{\phi\phi}}\, d\theta d\phi\\
               &=& 4\pi \left(r_{\pm}^2+2qr_{\pm}+a^2\right)\\
               &=& 8\pi {\cal M} r_{\pm}~.\label{hr2}
\end{eqnarray}
The horizon area of the BH is a 2D surface and it is constant as pointed out by the
Hawking~\cite{bcw73}. Also it never decreases. The other thermodynamic quantites like 
the surface gravity, the angular velocity and the Hawking temperature 
of (${\cal H}^{+}$)~\cite{as92} and of (${\cal H}^{-}$)~\cite{epjc16} are 
\begin{eqnarray}
{\Omega}_{\pm} &=& \frac{a}{r_{\pm}^2+2qr_{\pm}+a^2} =\frac{a}{2{\cal M}r_{\pm}} ~. \label{omega}\\
T_{\pm}&=& \frac{r_{\pm}-r_{\mp}}{4\pi\left(r_{\pm}^2+2qr_{\pm}+a^2 \right)} 
=\frac{r_{\pm}-r_{\mp}}{8\pi {\cal M} r_{\pm}},  \,\,\,  T_{+} > T_{-}  ~.\label{tmKN}
\end{eqnarray}
The horizon Killing vector field may be defined for ${\cal H}^\pm$ 
\begin{eqnarray}
{\chi_{\pm}}^{a} &=& (\partial_{t})^a +\Omega_{\pm} (\partial_{\phi})^a~.\label{hkv}
\end{eqnarray}

\section{\label{ksen1} The Surface energy, The Rotational energy and The Electromagnetic energy 
for Kerr-Sen BH:}
Larry Smarr~\cite{smarr73a,smarr73b} first derived the mass formula for 
charged spinning BH i.e. Kerr-Newman BH which could be obtained by integrating the term 
by term of the mass differential which consists of three terms interpreted as the \emph{surface energy}, 
\emph{rotational energy}, and \emph{electromagnetic energy}~\footnote{For recent calculation of these 
energies for the NUT class of BHs one must see the Ref.~\cite{pp21}} respectively. Here 
we \emph{extend this computation for a rotating charged BH in heterotic string theory}. The squared mass 
for Kerr-Sen BH derived in~\cite{epjc16} is 
\begin{eqnarray}
{\cal M}^2 &=& \frac{{\cal A}_{\pm}}{16\pi}+\frac{4\pi J^2}{{\cal A}_{\pm}} +\frac{Q^2}{2} ~.\label{mas}
\end{eqnarray}
The mass differential for both the horizons is 
\begin{eqnarray}
d{\cal M} &=&\Upsilon_{\pm} d{\cal A}_{\pm} + \Omega_{\pm} dJ +\Phi_{\pm} dQ ~. \label{dm}
\end{eqnarray}
where
\begin{eqnarray}
{\Upsilon}_{\pm} &=& \frac{\partial {\cal M}}{\partial {\cal A}_{\pm}}= \frac{1}{{\cal M}} 
\left(\frac{1}{32 \pi}-\frac{2\pi J^2}{{\cal A}_{\pm}^2}\right)\nonumber \\
\Omega_{\pm} &=& \frac{\partial {\cal M}}{\partial J}=\frac{4\pi J}{{\cal M}{\cal A}_{\pm}}
=\frac{a}{2{\cal M}r_{\pm}} \nonumber\\
\Phi_{\pm} &=& \frac{\partial {\cal M}}{\partial Q}=\frac{Q}{2{\cal M}}   ~. \label{invar}
\end{eqnarray}
and 
\begin{eqnarray}
{\Upsilon}_{\pm} &=& \mbox{The effective surface tension for ${\cal H}^{\pm}$}
\nonumber \\
\Omega_{\pm} &=&  \mbox{The angular velocity for ${\cal H}^\pm$} \nonumber \\
\Phi_{\pm} &=& \mbox{The electromagnetic potentials for ${\cal H}^\pm$}\nonumber
\end{eqnarray}

Remarkably, the parameters ${\Upsilon}_{\pm}$, $\Omega_{\pm}$ and $\Phi_{\pm}$ should be defined 
and are constant on the horizons for any stationary axisymmetric BH. Since $d{\cal M}$ is a 
perfect differential hence one could choose any convenient path of integration in $(A, J, Q)$
space. In particular, one could choose a path that will be defined for a charged 
BH in heterotic string theory. It has three energy components: 
the surface energy ${\cal E}_{s}^{\pm}$ defined in~\cite{epjc16}~[Eq.(76)]  
for ${\cal H}^{\pm}$ as
\begin{eqnarray}
{\cal E}_{s}^{\pm} &=& \int_{0}^{{\cal A}_{\pm}}\Upsilon_{\pm}(\tilde{{\cal A}_{\pm}}
, 0 ,0)\, d\tilde{{\cal A}_{\pm}}; ~ \label{se}
\end{eqnarray}
the rotational energy  for ${\cal H}^{\pm}$~[Eq.(77) in Ref.~\cite{epjc16} ] is
\begin{eqnarray}
{\cal E}_{r}^{\pm} &=& \int_{0}^{J} \Omega_{\pm} ({\cal A}_{\pm}
, \tilde{J} ,0)\, d\tilde{J},\,\,  \mbox{${\cal A}_{\pm}$ fixed}; ~ \label{re}
\end{eqnarray}
and the electromagnetic energy~[Eq.(78) in Ref.~\cite{epjc16}] for ${\cal H}^{\pm}$ is
\begin{eqnarray}
{\cal E}_{em}^{\pm} &=& \int_{0}^{Q} \Phi_{\pm} ({\cal A}_{\pm}
, J, \tilde{Q})\, d\tilde{Q},\,\, \mbox{${\cal A}_{\pm}$, $J$  fixed}; ~ \label{re1}
\end{eqnarray}

The mass parameter or BH mass parameter in Eq.~(\ref{mas}) may be  rewritten as
\begin{eqnarray}
{\cal M}\,({\cal A}_{\pm}, J, Q) &=& \sqrt{\frac{{\cal A}_{\pm}}{16\pi}
+\frac{4\pi J^2}{{\cal A}_{\pm}}+\frac{Q^2}{2}} ~.\label{mas1}
\end{eqnarray}

The above energy integrals should be directly computed using the variational definitions 
which is defined in Eq.~({\ref{invar}}).
First, we would like to compute the surface energy of ${\cal H}^{\pm}$ 
\begin{eqnarray}
{\cal E}_{s}^{\pm} &=& \sqrt{\frac{{\cal A}_{\pm}}{16\pi}} ~ \label{tn18}
\end{eqnarray}
Next we evaluate the rotational energy of ${\cal H}^{\pm}$ as
\begin{eqnarray}
{\cal E}_{r}^{\pm} &=& \sqrt{\frac{{\cal A}_{\pm}}{16\pi}+\frac{4\pi\,J^2}{{\cal A}_{\pm}}}
                       -\sqrt{\frac{{\cal A}_{\pm}}{16\pi}}.~ \label{tn19}
\end{eqnarray}
and finally we calculate the electromagnetic energy  of ${\cal H}^{\pm}$  as 
\begin{eqnarray}
{\cal E}_{em}^{\pm} &=& \sqrt{\frac{{\cal A}_{\pm}}{16\pi}+\frac{4\pi\,J^2}{{\cal A}_{\pm}}+\frac{Q^2}{2}} 
     -\sqrt{\frac{{\cal A}_{\pm}}{16\pi}+\frac{4\pi\,J^2}{{\cal A}_{\pm}}}.~ \label{tn20}
\end{eqnarray}
Now we compute the sum of three energies 
\begin{eqnarray}
{\cal E}_{s}^{\pm}+{\cal E}_{r}^{\pm}+{\cal E}_{em}^{\pm} &=&  
\sqrt{\frac{{\cal A}_{\pm}}{16\pi}+\frac{4\pi J^2}{{\cal A}_{\pm}}+\frac{Q^2}{2}}.
~ \label{tn21}
\end{eqnarray}
Using Eq.~({\ref{mas1}}), we can rewrite the above equation as
\begin{eqnarray}
{\cal E}_{s}^{\pm}+{\cal E}_{r}^{\pm}+{\cal E}_{em}^{\pm} 
&=& {\cal M}\,({\cal A}_{\pm}, J, Q) .~ \label{tn22}
\end{eqnarray}
Remarkably, the mass can be expressed as the sum of three energies, 
namely the surface energy of ${\cal H}^{\pm}$, the rotational energy of ${\cal H}^{\pm}$ 
and the electromagnetic energy of ${\cal H}^{\pm}$. This is one of the \emph{key} results
of the paper.  

\section{\label{ksen3} The Scale Parameter and Distortion Parameter for Kerr-Sen BH:}
Like Kerr-Newman BH~\cite{smarr73c}, it is convenient to study geometry intrinsic to 
the surface of charged BH in heterotic string theory. Therefore we would like to introduce
the scale parameter ~$(\zeta_{\pm})$, the distortion parameter~($\xi_{\pm}$) and a 
new parameter~$(\sigma_{\pm})$ which corresponds to the BH area~(${\cal A}_{\pm}$), 
angular momentum ~$(J)$ and charge~($Q$). It is helpful to study the 
\emph{intrinsic geometry} of both ${\cal H}^{\pm}$. They should  be  defined as 
\begin{eqnarray}
\zeta_{\pm} &=& \sqrt{\frac{{\cal A}_{\pm}}{4\pi}}, \,\, 
\xi_{\pm} = \frac{a}{\zeta_{\pm}},\,\, 
\sigma_{\pm} = \frac{Q}{\zeta_{\pm}} ~. \label{epsil}
\end{eqnarray}
In terms of these parameters, the three energy can be written  as 
\begin{eqnarray}
{\cal E}_{s}^{\pm} &=&  \frac{\zeta_{\pm} }{2}
\end{eqnarray}
\begin{eqnarray}
{\cal E}_{r}^{\pm} &=& \frac{\zeta_{\pm} }{2} \left[\frac{1}{\sqrt{1-\xi_{0}^2}}-1 \right]
\end{eqnarray}
where 
$$
\xi_{0}=\xi({\cal A}_{\pm}, J, Q=0)
$$
and 
\begin{eqnarray}
{\cal E}_{em}^{\pm} &=&  \frac{\zeta_{\pm} }{2} 
\left[\sqrt{\frac{1+2\,\sigma_{\pm}^2}{1-\xi_{\pm}^2}}-\frac{1}{\sqrt{1-\xi_{0}^2}} \right]~. \label{em1.1}
\end{eqnarray}
The integrated mass formula is then given by 
\begin{eqnarray}
{\cal M} &=&  \frac{\zeta_{\pm} }{2}\sqrt{\frac{1+2\,\sigma_{\pm}^2}{1-\xi_{\pm}^2}}~. \label{e1.2}
\end{eqnarray}
This is an another \emph{key} observation of this work. 

\section{\label{ksen4} Intrinsic Surface Geometry for Kerr-Sen BH}
In the previous section, we have described the geometry intrinsic to the surface of a 
Kerr-Sen BH in terms of the scale parameter and distortion parameter. In this section 
we will write  the surface geometry of Kerr-Sen BH described by the 2-metric
\begin{eqnarray}
d{\cal S}^2 &=&  (\omega_{\theta})^2+(\omega_{\phi})^2~. \label{g1.1}
\end{eqnarray}
where 
\begin{eqnarray}
\omega_{\theta} &=& \zeta_{\pm} \sqrt{1-\xi_{\pm}^2 \sin^2\theta}\,d\theta,  ~ \label{g1.2}\\
\omega_{\phi}   &=& \frac{\zeta_{\pm} \sin\theta\,d\phi}{\sqrt{1-\xi_{\pm}^2 \sin^2\theta}}
~.\label{g1.3}
\end{eqnarray}
Alternatively, a set of values for $\eta_{\pm}$ and $\beta_{\pm}$ determine only 
the spin parameter uniquely:
\begin{eqnarray}
a &=& \xi_{\pm} \zeta_{\pm}, ~. \label{a1}
\end{eqnarray}
On the other hand an algebraic expression relates the mass parameter and the charge parameter:
\begin{eqnarray}
{\cal M} &=& \frac{\zeta_{\pm}}{2} \left(1-\xi_{\pm}^2\right)^{-\frac{1}{2}} \left(1+2\frac{Q^2}{\zeta_{\pm}^2}\right)^{1/2}
~. \label{m1}
\end{eqnarray}
It should be noted that the angular momentum parameter relates the charge parameter as 
\begin{eqnarray}
J &=& \frac{\zeta_{\pm}^2}{2} \xi_{\pm}\left(1-\xi_{\pm}^2\right)^{-\frac{1}{2}} 
\left(1+2\frac{Q^2}{\zeta_{\pm}^2}\right)^{1/2}~. \label{j1}
\end{eqnarray}
Eq.~(\ref{a1}) and Eq.~(\ref{m1}) determine the \emph{identical intrinsic surface geometry} of 
whole rotating charged BH in heterotic string theory. Since there exist the horizons $r_{\pm}$ 
which are real therefore there must be a condition 
\begin{eqnarray}
2 {\cal M}^2 \geq 2J+Q^2 ~. \label{k1.2}
\end{eqnarray}
which gives the range of charge and total mass-energy for the Kerr-Sen BH. Accordingly, one finds 
the following \emph{upper bound on charge and mass}:
\begin{eqnarray}
0\leq Q \leq \zeta_{\pm}\sqrt{\frac{1}{2\xi_{\pm}^2}-1},\\
\frac{\zeta_{\pm} }{2}\frac{1}{\sqrt{1-\xi_{\pm}^2}}  \leq {\cal M} \leq \frac{\zeta_{\pm}}{2\xi_{\pm}} 
\end{eqnarray}
When the BH satisfy the \emph{equality} in Eq.~(\ref{k1.2}), the BH is said to be an \emph{extreme} 
Kerr-Sen BH. {It must be noted that the maximum value of $\xi_{\pm}$ should be 
obtained when the value of the charge parameter becomes zero; }
\begin{eqnarray}
(\xi_{\pm})_{max}=\frac{1}{\sqrt{2}}\backsimeq 0.707
\end{eqnarray}
{When it exceeds the above limit~[Eq.~(\ref{k1.2})] the situation is called a naked singularity}. 

\section{\label{ksen5} Horizon, Angular Velocity,  Surface Area and  Irreducible mass in 
terms of Scale parameter and Distortion Parameter}
The horizons of ${\cal H}^{\pm}$ in terms of the \emph{scale parameter and distortion parameter} become 
\begin{eqnarray}
r_{\pm} &=& \zeta_{\pm} \frac{\sqrt{1-\xi_{\pm}^2}}{\sqrt{1+2\,\sigma_{\pm}^2}} ~. \label{5.1}
\end{eqnarray}
Similarly, the angular velocity of ${\cal H}^{\pm}$ in terms of these parameters can be written as 
\begin{eqnarray}
 \Omega_{\pm} &=& \frac{\xi_{\pm}}{\zeta_{\pm}}, ~. \label{5.0}
\end{eqnarray}
and the surface area of ${\cal H}^{\pm}$ becomes 
\begin{eqnarray}
{\cal A}_{\pm} &=& \int \omega_{\theta}\wedge\omega_{\phi} =4\pi\zeta_{\pm}^2 ~.\label{5.2}
\end{eqnarray}
where $\wedge$ denotes the wedge product of differential forms.
Interestingly the surface area of 2-surfaces  depends only on the \emph{scale parameter}. Now  we 
should define the Christodoulou's irreducible mass of ${\cal H}^{\pm}$ in terms of  
the \emph{scale parameter} as 
\begin{eqnarray}
{\cal M}_{irr,\pm} &=& \frac{\zeta_{\pm}}{2} ~. \label{5.3}
\end{eqnarray}
It immediately indicates that 
\begin{eqnarray}
{\cal A}_{\pm} &=& 16 \pi{\cal M} _{irr, \pm}^2 ~.\label{5.4}
\end{eqnarray}
Hence there is no physical process that can reduce the irreducible mass of the BH. This is the 
Christodoulou's argument of Hawking's theorem and which is applicable for a charged BH in 
heterotic string theory. Since the irreducible mass parameter~(${\cal M}_{irr}$) or 
the scale parameter $\zeta_{\pm}$, therefore the  surface area ~(${\cal A}_{\pm}$) remains 
constant in these processes so it is called as \emph{reversible transformations}.

\section{\label{ksen6} Gross Measure of the Surface Deformation}
\begin{figure}
\begin{center}
\subfigure[]{
\includegraphics[width=3in,angle=0]{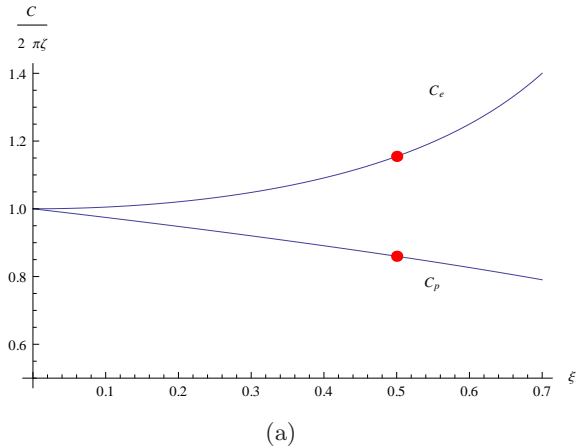}} 
\caption{The plot denotes the variation of the proper equatorial circumference and the proper 
polar circumference of Kerr-Sen BH with respect to the distortion parameter. 
The equatorial circumference~($C_{e}$) is increasing when the distortion  parameter is increasing. 
While the polar circumference is decreasing when the distortion parameter is increasing. 
{One can observe the bulging of the equator and flattening of the poles 
caused by the rotation. It should be remembered that the maximum value $\xi_{\pm}$ should 
obtain is determined by the value of $Q$. The red point on the curves implies the extremal 
Kerr BH when $Q=\zeta_{\pm}$~($\xi_{\pm}=\frac{1}{2}$).}}
\label{shift} 
\end{center}
\end{figure}

To determine the gross measure of the surface deformation one can compare the equatorial  
circumference $C_{e}$ and the polar circumference $C_{p}$, which are defined as the proper 
length of the curves $\theta=\frac{\pi}{2}$ and $\phi=0$, respectively. First we calculate 
the proper equatorial circumference that is 
\begin{eqnarray}
C_{e} &=& \int \omega_{\phi}=\frac{2\pi \zeta_{\pm}}{\sqrt{1-\xi_{\pm}^2}}~.\label{ce}
\end{eqnarray}
and the proper polar circumference is defined as 
\begin{eqnarray}
C_{p} &=& \int \omega_{\theta}=4\zeta_{\pm} E(\xi_{\pm}) 
\end{eqnarray}
which is a complete elliptical integral of the second kind. {These 
circumferences are invariant quantity because the curves are geodesics of the 2-metric. 
In Fig.~\ref{shift}, we have drawn a graph of $C_{e}$ and $C_{p}$ with $\xi$ in a reversible 
transformation. From the graph, it shows that as $\xi$ increases the parameter $C_{e}$ increases 
and $C_{p}$ decreases.}

Another quantity that is a measure of the oblateness or prolateness of the BH should be 
determined by the  equatorial circumference and polar circumferences i.e. 
$C_{e}$ and $C_{p}$. The dimensionless number 
\begin{eqnarray}
\lambda &=& 1-\frac{C_{p}}{C_{e}}=1-\frac{2}{\pi} \sqrt{1-\xi_{\pm}^2}\,\, E(\xi_{\pm}) 
\end{eqnarray}
which determines the degree of oblateness~(or prolateness). This parameter tells us how 
oblate the BH surface is becoming. {It is plotted with respect to the 
distortion parameter in Fig. 2. The implication of proper equatorial 
circumference [Eq. (\ref{ce})] indicates that one could consider the distortion parameter $\xi_{\pm}$ 
as being the product of the angular velocity  of the BH with its effective radius. From the 
definition of angular velocity defined in Eq.~(\ref{5.0}) one obtains the distortion 
parameter as 
\begin{equation}
\xi_{\pm}=\Omega_{\pm} \zeta_{\pm} 
\end{equation}
Hence following the definition of Christodoulou~\cite{cr70}, the equatorial surface velocity 
of a BH can be defined as 
\begin{equation}
v=\sqrt{g_{\phi\phi}}\Omega_{\pm} =\frac{\xi_{\pm}}{\sqrt{1-\xi_{\pm}^2}} 
\end{equation}
The significance of this definition is that when the extreme Kerr hole is approched i.e. 
$\xi_{\pm}\rightarrow \frac{1}{\sqrt{2}}$ then the velocity tends to the velocity of 
light  i.e. $v\rightarrow 1$. Now it is clear to us that why the BH surface disappears 
altogeather for $\xi_{\pm}>(\xi_{\pm})_{max}$. }

\begin{figure}
\begin{center}
\subfigure[]{
\includegraphics[width=3in,angle=0]{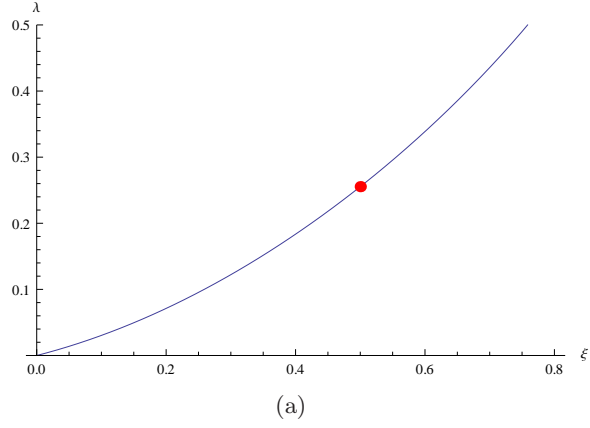}} 
\caption{This plot shows the variation of the surface deformation on a BH 
caused by rotation. The {parameter} $\lambda$ determines the 
`oblateness' of BH surface.}
\label{shift1} 
\end{center}
\end{figure}
The Gaussian curvature is an important quantity in differential geometry which measures 
intrinsic property of space geometry. Also it is independent of the coordinate system used 
to narrate it. It is denoted by the symbol $K$.  It should be noted that the Gaussian curvature 
for a sphere is $K=\frac{1}{r^2}>0$, for a plane $K=0$ and for a pseudosphere $K=-\frac{1}{r^2}<0$.
The detailed of mathematical derivation of Gaussian curvature should be derived in Appendix-A. 

\section{\label{ksen8} How much rotational energy can be extracted from Kerr-Sen BH?}
In a seminal work~\cite{penrose73}  Penrose \& Floyd first showed how one can extract the rotational 
energy from a spinning~(Kerr) BH. This is possible only due to the presence of the finite region
of space~(ergosphere)~\footnote{Note that the ergosphere coincides with $r_{+}$ only at the 
poles $\theta=0$ and $\theta=\pi$} which is located  between the event horizon~($r_{+}$) and stationary limit 
surface~($r_{ergo}$) i.e. $r_{+}<r<r_{ergo}(\theta)$. The finite region has an 
important effect because it allows to extract the rotational energy from the BH. 
Here, we will derive this \emph{rotational  energy  for a charged rotating BH in 
heterotic string theory } by  using Eq.~(\ref{re}). We find the rotational energy 
for Kerr-Sen BH as 
\begin{eqnarray}
{\cal E}_{r}^{+} &=& \int_{0}^{J} \Omega_{+} ({\cal A}_{+}
, \tilde{J} ,0)\, d\tilde{J},\,\,  \mbox{${\cal A}_{+}$ fixed}; ~ \label{ere}
\end{eqnarray}
Now the  above calculation yields 
$$
{\cal E}_{r}^{KS} = \sqrt{\frac{{\cal M}}{2}\left[\left({\cal M}-\frac{Q^2}{2{\cal M}}\right)
+\sqrt{\left({\cal M}-\frac{Q^2}{2{\cal M}}\right)^{2}-a^2} \right]}\times
$$
\begin{eqnarray}
\left[\sqrt{\frac{2\left({\cal M}-\frac{Q^2}{2{\cal M}}\right)}
{\left[\left({\cal M}-\frac{Q^2}{2{\cal M}}\right)
+\sqrt{\left({\cal M}-\frac{Q^2}{2{\cal M}}\right)^{2}-a^2} \right]}}-1\right]~. \label{3.6}              
\end{eqnarray}
It indicates that \emph{this is the maximum amount of rotational energy we can extract from the rotating 
charged BH in heterotic string theory}.  It also implies that the quantity ${\cal E}_{r}^{KS}$ depends on 
both the charge and spin parameter. 

Now we will compare the above result with the Kerr-Newman BH. For KN BH, the 
rotational energy yields 
$$
{\cal E}_{r}^{KN} =\frac{\sqrt{2{\cal M}\left({\cal M}+\sqrt{{\cal M}^2-a^2-Q^2}-Q^2\right)}}{2} \times
$$
\begin{eqnarray}
\left[\sqrt{1+\frac{4a^2{\cal M}^2}{\left[2{\cal M}\left({\cal M}+\sqrt{{\cal M}^2-a^2-Q^2}-Q^2\right)\right]^2}}-1 \right] 
\end{eqnarray}

It should be noted that when the charge parameter goes to zero value we get the value of rotational 
energy for Kerr BH~\cite{sc}  
\begin{eqnarray}
{\cal E}_{r}^{Kerr} &=& \sqrt{\frac{{\cal M}}{2}\left({\cal M}+\sqrt{{\cal M}^{2}-a^2} \right)}
\left[\sqrt{\frac{2{\cal M}} {\left({\cal M}+\sqrt{{\cal M}^{2}-a^2} \right)}}-1\right] 
~. \label{3.8}              
\end{eqnarray}
Firstly taking the extremal limit of Kerr BH~($a={\cal M}$), we find 
${\cal E}_{r}^{XKerr}=\left(\frac{\sqrt{2}-1}{\sqrt{2}}\right) {\cal M}$ i.e. we can get  
approximately 29 percentage of its total energy.  

Now taking the \emph{extremal limit of Kerr-Sen BH}, we get 
\begin{eqnarray}
{\cal E}_{r}^{XKS} &=& \left(\frac{\sqrt{2}-1}{2}\right)\sqrt{2{\cal M}^2-Q^2}
=\left(\frac{\sqrt{2}-1}{\sqrt{2}}\right) \sqrt{J} ~.\label{4.0}             
\end{eqnarray}
It follows that the energy value strictly depends on charge parameter or 
the  angular momentum parameter. This is the \emph{last key} result of this work. 
To the best of our knowledge, this is the \emph{first time} we have examined the 
energy extraction process for a BH in heterotic string theory. 

Finally for our record, taking the extremal limit of Kerr-Newman BH~($a^2+Q^2={\cal M}^2$) 
then we find  
\begin{eqnarray}
\frac{{\cal E}_{r}^{XKN}}{{\cal M}} &=& \frac{\sqrt{1+\frac{J^2}{{\cal M}^4}}}{2}
\left[\sqrt{1+\frac{\frac{4J^2}{{\cal M}^4}}{\left(1+\frac{J^2}{{\cal M}^4}\right)^2}}-1 \right]  
~. \label{4.1}              
\end{eqnarray}


In the limit $J={\cal M}^2$, the rotational energy  is indeed equal to extremal Kerr BH.



\section{\label{dis} Discussion}
In this paper we have considered three parameter charged rotating BH solutions in the low energy limit 
of effective field theory in heterotic string theory. We computed the \emph{surface energy, the 
rotational energy and the electromagnetic energy} for a spinning charged BH in heterotic string 
theory having the event horizon and the Cauchy horizon. Remarkably, we showed that the sum of 
three energies is equal to the mass parameter i.e.  
$$
{\cal E}_{s}^{\pm}+{\cal E}_{r}^{\pm}+{\cal E}_{em}^{\pm}={\cal M}
$$
Moreover, in terms of the \emph{ scale parameter, the distortion parameter and 
a new parameter~$(\sigma_{\pm})$} which corresponds to the surface area of the BH, the 
angular momentum  and the charge parameter. We proved that the  mass parameter in a 
compact form like
$$
{\cal E}_{s}^{\pm}+{\cal E}_{r}^{\pm}+{\cal E}_{em}^{\pm}=
{\cal M} =\frac{\zeta_{\pm} }{2}\sqrt{\frac{1+2\,\sigma_{\pm}^2}{1-\xi_{\pm}^2}}
$$
which is valid for ${\cal H}^{\pm}$. Furthermore, we studied the intrinsic surface 
geometry of the BH. We determined the upper bound of mass and charge parameter in 
terms of scale parameter and distortion parameter 
\begin{eqnarray}
0\leq Q \leq \zeta_{\pm}\sqrt{\frac{1}{2\xi_{\pm}^2}-1},\\
\frac{\zeta_{\pm} }{2}\frac{1}{\sqrt{1-\xi_{\pm}^2}}  \leq {\cal M} \leq \frac{\zeta_{\pm}}{2\xi_{\pm}} 
\end{eqnarray}
We also determined the horizon radius, angular velocity, surface area and irreducible mass 
in terms of scale paprameter and distortion parameter. We computed the equatorial circumference 
and polar circumference for Kerr-Sen BH which are gross measure of BH surface deformation. 
We proved that when the spinning rate of the BH increases the equatorial circumference 
increases but polar circumference decreases. 

Next we calculated the Gaussian curvature for Kerr-Sen BH. We proved that there exists two 
class of geometry separated by $\xi_{\pm}=\frac{1}{2}$ for this type of BH. In the regime  
$\frac{1}{2}<\xi_{\pm}\leq \frac{1}{\sqrt{2}}$, the Gaussian curvature is negative and 
there exist \emph{two polar caps} on the surface. While for $\xi_{\pm}<\frac{1}{2}$, 
the Gaussian curvature is positive and the surface would be an oblate deformed sphere.
We also derived the topology and the Euler  number  by using Gauss-Bonnet theorem.

Finally, we  computed the value of rotational energy that should be extracted from 
the spinning charged BH in heterotic string theory via famous Penrose process. 
The maximum value of extracted rotational energy for \emph{extremal} Kerr-Sen BH 
is found to be 
$$
{\cal E}_{r}^{+}=
\left(\sqrt{2}-1\right)\sqrt{\frac{J}{2}}
$$

\section{Appendix-A}
To compute the Gaussian curvature we have to write the metric in the following form by using 
Eq. (\ref{g1.1})
\begin{eqnarray}
d{\cal S}^2 &=& \zeta_{\pm}^2 \left[\frac{dy^2}{f(y)}+f(y)\,d\phi^2 \right]~. \label{7.1}
\end{eqnarray}
where  
\begin{eqnarray}
f(y) &=& \frac{1-y^2}{1-\xi_{\pm}^2(1-y^2)}~. \label{7.2}
\end{eqnarray}
and the value of $y=\cos\theta$. We know that the Gaussian curvature~\cite{smarr73c} is defined for the 
above metric as 
\begin{eqnarray}
K \equiv K(y) = - \frac{f''(y)}{2\zeta_{\pm}^2}~. \label{7.3}
\end{eqnarray}
For our metric, the Gaussian curvature of ${\cal H}^{\pm}$ is calculated to be 
\begin{eqnarray}
K_{\pm}({\cal M},a,q,\theta) &=& \frac{\left(r_{\pm}^2+2qr_{\pm}+a^2 \right)
\left(r_{\pm}^2+2qr_{\pm}-3a^2\cos^2\theta\right)}
{\left(r_{\pm}^2+2qr_{\pm}+a^2\cos^2\theta\right)^3}~. \label{7.4}
\end{eqnarray}
In terms of scale parameter and distortion parameter, the Gaussian curvature 
of ${\cal H}^{\pm}$ becomes 
\begin{eqnarray}
K_{\pm}(\zeta_{\pm},\xi_{\pm},\theta) &=& \frac{\left[1-\xi_{\pm}^2\left(1+3\cos^2\theta\right)\right]}
{\zeta_{\pm}^2 \left(1-\xi_{\pm}^2 \sin^2\theta\right)^3}~. \label{7.5}
\end{eqnarray}
When the rotation is off i.e. $a=\xi_{\pm}=0$, we will get the spherically symmetric situation and 
in this situation the Gaussian curvature becomes
\begin{eqnarray}
K_{\pm}(\zeta_{\pm},0,\theta) &=& \frac{1}{\zeta_{\pm}^2}=\frac{1}{r_{\pm}^2+2qr_{\pm}}
=K_{\pm}({\cal M},0,q,\theta)~. \label{7.51}
\end{eqnarray}
If one could take the value of charge parameter becomes zero then we would get spherically 
symmetric Schwarzschild BH:
\begin{eqnarray}
K_{\pm}(\zeta_{\pm},0,\theta) &=& \frac{1}{\zeta_{+}^2}=\frac{1}{r_{+}^2}
=K_{\pm}({\cal M},0,0,\theta)~. \label{7.52}
\end{eqnarray}
It implies that the Gaussian curvature looks like a sphere. This is the limiting situation. 
For generalized case the Gaussian curvature is a function of polar angle $\theta$ when the 
the BH is rotating i.e. ($\xi_{\pm} \neq 0$). 

Case-I: When $\xi_{\pm}=\frac{1}{2}$, the Gaussian curvature becomes zero at the 
poles~($\theta=0$). Alternatively this occurs  when $a=\frac{\sqrt{3{\cal M}^2-2Q^2}}{2}$

Case-II: When $\frac{1}{2}<\xi_{\pm}\leq \frac{1}{\sqrt{2}}$, the Gaussian curvature becomes 
negative and there exist two polar caps on the surface. Like Kerr-Newman BH, there exist 
\emph{two geometrically distinct classes} of charged rotating BH in heterotic string theory when the 
value of distortion parameter satisfied the following criterion:
$$
\xi_{\pm} \gtrless \frac{1}{2}
$$
Say for example when we choose the criterion $\xi_{\pm}>\frac{1}{2}$, the Gaussian curvature 
is computed to be
\begin{eqnarray}
K_{\pm}(\zeta_{\pm},\frac{3}{5},\theta) &=& \frac{625}{\zeta_{\pm}^2}\frac{(16-27\cos^2\theta)}{(25-9\sin^2\theta)^3}. 
\end{eqnarray}
Consequently at the pole it becomes 
\begin{eqnarray}
K_{\pm}(\zeta_{\pm},\frac{3}{5},0) &=&- \frac{11}{25 \zeta_{\pm}^2}<0. 
\end{eqnarray}
This class of BH geometry having the Gaussian curvature is negative. That's why the surfaces are likely 
to be a pseudosphere.

On the other hand if we choose the criterion $\xi_{\pm}<\frac{1}{2}$, the Gaussian curvature 
is computed to be
\begin{eqnarray}
K_{\pm}(\zeta_{\pm},\frac{2}{5},\theta) &=& \frac{625}{\zeta_{\pm}^2}\frac{(21-12\cos^2\theta)}{(25-4\sin^2\theta)^3}. 
\end{eqnarray}
Consequently at the pole it becomes 
\begin{eqnarray}
K_{\pm}(\zeta_{\pm},\frac{2}{5},0) &=& \frac{9}{25 \zeta_{\pm}^2}>0. 
\end{eqnarray}
This class of BH geometry having the Gaussian curvature is positive everywhere on the surface. 
The  surfaces are likely to be a oblate deformed sphere. 

Note that as a special case at the equator, the curvature of ${\cal H}^{\pm}$ 
reduces to 
\begin{eqnarray}
K_{\pm}^{e} &=& \frac{\left(r_{\pm}^2+2qr_{\pm}+a^2 \right)}{\left(r_{\pm}^2+2qr_{\pm}\right)^2}
=\frac{2{\cal M}}{r_{\pm}\left(r_{\pm}+2q\right)^2}>0 ~. \label{7.6}
\end{eqnarray}

In summary, at the pole, there exist \emph{two polar caps} while at the \emph{equator} there does 
not exist such a polar cap.

\section{Appendix-B:  Topology, Euler number and Gauss-Bonnet Theorem}
Using the general formula of Gaussian curvature derived in Eq.~(\ref{7.5}), we can compute
the topology of the surface. To evaluate the topology of the surface we can apply the Gauss-Bonnet 
theorem which states that 
\begin{eqnarray}
\iint_{M} K_{\pm}\,\, \omega_{\theta}\wedge\omega_{\phi} &=& 2\pi \chi(M) ~. \label{7.10}
\end{eqnarray}
where $\chi$ is the Euler characteristic of the surface. For ${\cal H}^{\pm}$, we get 
\begin{eqnarray}
\iint_{M} K_{\pm}\,\, \omega_{\theta}\wedge\omega_{\phi} &=& 4\pi ~. \label{7.11} 
\end{eqnarray}
This implies that $\chi=2$. Like Kerr BH, the surface of the charged BH in heterotic string 
theory is \emph{topologically a 2-sphere}. 

\emph{Data Availability}\\
Data sharing not applicable to this article as no datasets were generated or analysed during the current study.

\bibliographystyle{model1-num-names}

\end{document}